\newcommand{\PRE}[1]{}       
\documentclass[aps,prl,twocolumn,preprintnumbers,superscriptaddress,
amsmath,amssymb,showpacs,nofootinbib]{revtex4}
\newcommand{\ssection}[1]{{\em #1.\ }}

\usepackage{enumerate}
\usepackage{color}
\usepackage[pdftex]{graphicx}
\usepackage[colorlinks=true,citecolor=blue,urlcolor=blue]{hyperref}
\graphicspath{ {../Figures/} }

\newcommand{\mplanck}{M_{\text{Pl}}}

\newcommand{\OmegaDM}{\Omega_{\text{DM}}}

\newcommand{\kev}{\text{keV}}
\newcommand{\mev}{\text{MeV}}
\newcommand{\gev}{\text{GeV}}
\newcommand{\tev}{\text{TeV}}

\newcommand{\cm}{\text{cm}}

\newcommand{\g}{\text{g}}
\newcommand{\s}{\text{s}}

\newcommand{\kpc}{\text{kpc}}

\newcommand{\Eqref}[1]{Equation~(\ref{#1})}
\renewcommand{\eqref}[1]{Eq.~(\ref{#1})}
\newcommand{\eqsref}[2]{Eqs.~(\ref{#1}) and (\ref{#2})}

\newcommand{\figref}[1]{Fig.~\ref{fig:#1}}

\newcommand{\tableref}[1]{Table~\ref{table:#1}}

\newcommand{\gb}{G}                            
\newcommand{\gbino}{\tilde{G}}                 
\newcommand{\gbinoX}{\tilde{G}^*}              

\begin{document}

\preprint{UCI-TR-2014-06, CERN-PH-TH-2014-162, CALT-TH-2014-148}

\title{\PRE{\vspace*{1.5in}}
SIMPle Dark Matter: Self-Interactions and keV Lines
\PRE{\vspace*{.5in}}}

\author{Kimberly K.~Boddy}
\affiliation{Walter Burke Institute for Theoretical Physics,
  California Institute of Technology, Pasadena, CA 91125, USA
\PRE{\vspace*{.2in}}}

\author{Jonathan L.~Feng}
\affiliation{Department of Physics and Astronomy, University of
  California, Irvine, California 92697, USA
\PRE{\vspace*{.2in}}}
\affiliation{CERN Theory Division, CH-1211, Geneva 23, Switzerland
\PRE{\vspace*{.2in}}}

\author{Manoj Kaplinghat}
\affiliation{Department of Physics and Astronomy, University of
  California, Irvine, California 92697, USA
\PRE{\vspace*{.2in}}}

\author{Yael Shadmi}
\affiliation{Physics Department, Technion---Israel Institute of
  Technology, Haifa 32000, Israel
\PRE{\vspace*{.4in}}}

\author{Timothy M.~P.~Tait\PRE{\vspace*{.1in}}}
\affiliation{Department of Physics and Astronomy, University of
  California, Irvine, California 92697, USA
\PRE{\vspace*{.2in}}}


\begin{abstract}
\PRE{\vspace*{.2in}} We consider a simple supersymmetric hidden
sector: pure SU($N$) gauge theory.  Dark matter is made up of hidden
glueballinos with mass $m_X$ and hidden glueballs with mass near the
confinement scale $\Lambda$.  For $m_X \sim 1~\tev$ and $\Lambda \sim
100~\mev$, the glueballinos freeze out with the correct relic density
and self-interact through glueball exchange to resolve small-scale
structure puzzles.  An immediate consequence is that the glueballino
spectrum has a hyperfine splitting of order $\Lambda^2 / m_X \sim
10~\kev$.  We show that the radiative decays of the excited state can
explain the observed 3.5 keV X-ray line signal from clusters of
galaxies, Andromeda, and the Milky Way.
\end{abstract}


\maketitle

\ssection{Introduction} The field of particle dark matter is at an
interesting juncture.  Direct, indirect, and collider searches for
dark matter are improving rapidly, but have not yet yielded
unambiguous signals.  At the same time, the astrophysical evidence for
dark matter with $\OmegaDM h^2 = 0.1196 \pm 0.0031$~\cite{Ade:2013zuv}
remains as strong as ever, and there are now tantalizing astrophysical
indications that dark matter may be
self-interacting~\cite{Rocha:2012jg,Peter:2012jh,%
  Vogelsberger:2012ku,Zavala:2012us} or the source of an observed 3.5
keV X-ray line from galaxies and clusters of
galaxies~\cite{Bulbul:2014sua,Boyarsky:2014jta}.  Self-interactions
and the 3.5 keV line have each merited a great deal of attention,
although typically separately, without any attempt to relate them in a
simple framework.

Given the existing evidence for dark matter, a natural possibility is
that dark matter is in a hidden sector, composed of particles with no
standard model gauge interactions~\cite{Kobsarev:1966}.  In general,
hidden sectors are decoupled from most of particle physics, both in
terms of their theoretical motivations and their testable predictions.
In the framework of supersymmetry, however, hidden sectors may emerge
from more fundamental theories and contain particles that have the
desired thermal relic density through the WIMPless
miracle~\cite{Feng:2008ya,Feng:2008mu,Feng:2011ik,Feng:2011uf,Feng:2011in}.
Although much of our analysis below will be independent of
supersymmetry, the possibility of preserving this fundamental feature
of WIMPs is a significant virtue, and for concreteness, we will
consider hidden sectors with supersymmetry.

Here we consider the simplest possible UV-complete supersymmetric
hidden sector: a pure SU($N$) gauge theory. This sector introduces
only two new particles: gluons $g$ and gluinos $\tilde{g}$, which
hadronize into glueballs $G \equiv (gg)$ and glueballinos $\tilde{G}
\equiv (\tilde{g} g)$.  Throughout this work, references to color,
gluons, gluinos, and their composite states refer to the hidden
sector. For other work on strongly-interacting dark matter, see
Refs.~\cite{Kang:2006yd,Kribs:2009fy,Alves:2009nf,%
  Falkowski:2009yz,Lisanti:2009am,Alves:2010dd,Kumar:2011iy,%
  Higaki:2013vuv,Heikinheimo:2013fta,Bai:2013xga,Cline:2013zca,%
  Boddy:2014yra,Hochberg:2014dra,Cline:2014kaa,Juknevich:2009ji}.

We find that this simple hidden sector may explain all of the
above-noted astrophysical observations.  For gluino mass $m_X \sim
\tev$ and glueball mass near the confinement scale $\Lambda \sim
100~\mev$, glueballinos have both the correct relic density and
self-interaction cross section to resolve small-scale structure
puzzles~\cite{Boddy:2014yra}.  These considerations fix the
glueballino spectrum's hyperfine splitting $\Delta E = m_{\gbinoX} -
m_{\gbino} \sim \Lambda^2 / m_X \sim 10~\kev$. Introducing connector
fields that couple the hidden and visible sectors, we find that
radiative decays $\gbinoX \to \gbino \gamma$ may have the energy and
flux required to explain the observed X-ray line signals in both
``short lifetime'' and ``long lifetime'' scenarios.

\ssection{Glueballino Relic Density and Reannihilation} The
supersymmetric pure SU($N$) hidden sector may be completely
characterized by the four parameters
\begin{equation}
m_X, \Lambda, N, \xi_f \ ,
\end{equation}
which are the gluino mass, the confinement scale, the number of
colors, and the ratio of hidden to visible sector temperatures at
gluino freezeout, respectively.  In terms of these parameters, the
fine-structure constant at the scale $m_X$ is given by the
renormalization group relation
\begin{equation}
  \alpha_X = \frac{6\pi}{11N \ln(m_X/\Lambda)} \ .
\end{equation}

Gluinos freeze out with relic density~\cite{Feng:2008mu}
\begin{equation}
  \Omega_{\tilde{g}} \approx \frac{s_0}{\rho_{c0}} 
\frac{\sqrt{g_*^\textrm{tot}}}{g_{*S}(T_f)}
  \frac{3.79\ x_f}{\mplanck \langle \sigma v \rangle } \ ,
\label{gbinodensity}
\end{equation}
where $s_0$ is the visible sector entropy today, $\rho_{c0}$ is the
critical density today, $g_{*S}(T_f)$ is the entropy effective number
of degrees of freedom in the visible sector at freezeout,
$g_*^\textrm{tot}(T_f) = g_*(T_f) + \xi_f^4\ 2(N^2-1)$, $\mplanck
\simeq 1.2 \times 10^{19}~\gev$, and $x_f \equiv m_X/T_f \approx 25
\xi_f$.  The gluinos annihilate through the $S$-wave process
$\tilde{g}\tilde{g} \to gg$ with cross section
\begin{equation}
  \langle \sigma v \rangle  = \frac{3}{8}\frac{N^2}{N^2-1}
  \frac{\pi\alpha_X^2}{m_X^2} \ .
\end{equation}

When the Universe cools to a temperature below $\Lambda$, the gluinos
and gluons hadronize into glueballinos and glueballs.  The
glueballinos then interact with an enhanced geometric cross section
$\sim \Lambda^{-2}$, which may initiate an era of reannihilation,
depleting the gluino relic density. However, the glueballinos
typically form in a state with high angular momentum
$L$~\cite{Kang:2006yd}. For the constituent gluinos to annihilate,
this bound state must first decay to a low-$L$ state by radiating
glueballs.  (Note that there are no hidden light pions or photons.)
Reannihilation therefore requires $\alpha_X^2 m_X \agt N_G \Lambda$,
where $N_G$ is the number of glueballs radiated. $N_G$ is at least 1.
More typically, it is the angular momentum of the bound state $N_G
\sim L \sim m_X v r \sim m_X \sqrt{\Lambda/m_X} \Lambda^{-1}$.  Below,
we will therefore exclude regions where $\alpha_X^2 \agt \sqrt{\Lambda
  / m_X}$, and re-annihilation may be significant. Note that this
constraint may be overly stringent, since glueballs will readily break
apart high-$L$ bound states.

\ssection{Glueball Relic Density and Cannibalization} After gluinos
freeze out, the gluons maintain thermal equilibrium.  Upon confinement,
the gluon energy density becomes the glueball energy
density~\cite{Boddy:2014yra}
\begin{equation}
\label{gbdensity}
  \Omega_\gb \approx \frac{s_0}{\rho_{c0}} 
\frac{2(N^2-1)}{g_{*S}(T_f)} \xi_f^3 \times
  \begin{cases}
    T_d^h & \! \! \textrm{for } T_d^h < \Lambda \\
    \Lambda & \! \! \textrm{otherwise ,}
  \end{cases} 
\end{equation}
where $T_d^h$ is the hidden sector temperature at the time of chemical
decoupling.  \Eqref{gbdensity} may be understood as follows: In the
absence of self-interactions, the glueballs decouple early, and the
relic density is simply the thermal number density multiplied by the
glueball mass $\sim \Lambda$.  With significant self-interactions,
however, the glueballs may remain in chemical equilibrium even after
the temperature drops below $\Lambda$ through, for example, $3 \to 2$
number-changing processes.  This depletion of glueball number is
referred to as cannibalization~\cite{Carlson:1992fn}.  Eventually, the
expansion of the Universe causes the glueballs to decouple at a
temperature $T_d^h$, and entropy and glueball number conservation
after decoupling imply that $\Omega_G \propto T_d^h$.  We have
numerically solved for the glueball density accounting for
cannibalization, following Ref.~\cite{Carlson:1992fn}, and find that
cannibalization reduces $\Omega_G$ by less than a factor of two in the
parameter range of interest.

It is also possible to eliminate the glueball relic density altogether
by postulating additional interactions with the visible sector.  For
example, before confinement, gluons may annihilate to sterile
neutrinos, which quickly decay to light visible sector particles
before they can annihilate back into gluons~\cite{Boddy:2014yra}.  We
will consider cases in which the glueball relic density is given by
\eqref{gbdensity}, and also those in which glueballs are effectively
absent.

\ssection{Self-Interactions} Discrepancies between simulations and
observations on small scales may be resolved if dark matter
self-interacts with cross section-to-mass ratio $\sigma / m \sim
1~\cm^2/\g \sim 1~\text{barn}/\gev$.  To determine the
self-interactions of glueballs and glueballinos, we follow the
analysis of Ref.~\cite{Boddy:2014yra}, which we summarize here.

For glueballs, we take the geometric cross section $\sigma_G =
4\pi/\Lambda^2$, which is of the desired size for $\Lambda \sim
100~\mev$.

Glueballino self-interactions are mediated by glueball exchange, which
we model as an attractive Yukawa potential $V(r) = - e^{-\Lambda r} /
r$.  The self-interaction cross section $\sigma_{\gbino}$ is $\langle
\sigma_T \rangle$, the transfer cross section $\sigma_T = \int d\Omega
(1 - \cos \theta) (d \sigma/d\Omega)$, averaged over Maxwell-Boltzmann
velocity distributions with characteristic velocities $v_0 = 40$, 100,
and 1000 km/s for dwarf galaxies, LSBs, and clusters, respectively.
For $m_X \sim \tev$, $\Lambda \sim 10~\mev$ gives the desired
self-interactions.

The case of mixed glueballino-glueball dark matter is much more
complicated.  As a very simple measure in this general case, we define
\begin{equation}
\sigma / m
= \frac{\sigma_{\gbino}}{m_X} \frac{\Omega_{\gbino}}{\OmegaDM}
+ \frac{\sigma_G}{\Lambda} \frac{\Omega_{G}}{\OmegaDM} \ ,
\label{sigmaoverm}
\end{equation}
which has the correct behavior in the limits of pure $\gbino$ or pure
$G$ dark matter and interpolates between them.

\ssection{Glueballino Hyperfine Structure and Transitions} There are
two $S$-wave glueballino states: the spin-$1/2$ ground state $\gbino$
and the spin-$3/2$ excited state $\gbinoX$~\cite{Chanowitz:1983ci}.
In the case of atomic hydrogen, the hyperfine splitting created by
the electromagnetic interactions is $\sim \alpha_{\text{EM}}^4 m_e^2
/m_p$.  In the present case, we expect the hidden chromomagnetic
interactions to yield hyperfine splittings 
\begin{equation}
  \Delta E = m_{\gbinoX} - m_{\gbino} = c_E \Lambda^2/m_X \ ,
  \label{Ephoton-keV}
\end{equation}
with $c_E \sim 1$ an order one coefficient that depends on the strong
dynamics.  Lattice results for the hyperfine splittings of $B$
mesons~\cite{Dowdall:2012ab} suggest $c_E \approx 5$ for those
systems.

In the absence of other interactions, the $\gbinoX$ state is stable.
To make contact with the X-ray observations, we introduce a connector
field with mass $m_C$ and both hidden color and visible
electromagnetic quantum numbers.  Dipole operators vanish, since the
gluino is a Majorana fermion.  But one-loop diagrams with virtual
heavy connectors induce at leading order the K\"{a}hler potential term
\begin{equation}
c_\tau \frac{1}{m_C^3}
\int \! d^4 \theta \, W_\alpha^h W^{\alpha} 
\, \overline{D}^{\dot{\alpha}} \overline{W}^h_{\dot{\alpha}}  
\, \overline{S} \ .
\end{equation}
This leads to the dimension-6 interaction
\begin{equation}
c_\tau \frac{m}{m_C^3}
\bar{\tilde{g}} \gamma^\mu D^\nu \tilde{g} \, F_{\mu \nu}
\ ,
\end{equation}
where $W^h$ ($W$) is the hidden (visible) gauge superfield, $F$ is the
visible electromagnetic field strength, $\langle S \rangle = \tilde{m}
\theta^2$ is a spurion representing the influence of supersymmetry
breaking, and $c_\tau$ is a dimensionless coefficient, which we
estimate to be $c_\tau \sim e \alpha_h / (4 \pi)$.  At the hadronic
level, this mediates the decay $\gbinoX \to \gbino \gamma$ with
lifetime
\begin{eqnarray}
\label{lifetime}
\lefteqn{\tau \sim
\frac{32 \pi^2}{\alpha_{\text{EM}} \alpha_h^2} 
\frac{m_C^6}{\tilde{m}^2 \Lambda^2 \Delta E^3} 
\sim 6.2 \! \times \! 10^{14}~\s} \\
&\times& \! \! \!
\left[ \frac{0.01}{\alpha_h} \right]^2 
\left[ \frac{m_C}{\tev} \right]^6
\left[ \frac{\tev}{\tilde{m}} \right]^2
\left[ \frac{100~\mev}{\Lambda} \right]^2
\left[ \frac{3.5~\kev}{\Delta E} \right]^3 \! , \nonumber
\end{eqnarray}
where $m_X \alt \tilde{m} \alt m_C$.  Decays to neutrinos or multiple
photons are also possible, but will be suppressed by additional powers
of $m_C$ and phase space.

\ssection{The 3.5 keV Line} The stacked XMM-Newton spectrum of 73
clusters of galaxies has revealed a weak X-ray line at $3.55-3.57$
keV~\cite{Bulbul:2014sua}. The line is also seen in the Perseus
cluster by both XMM-Newton and
Chandra~\cite{Bulbul:2014sua,Boyarsky:2014jta}, in the stacking of
Centaurus, Ophiuchus and Coma, and in the stacking of all clusters
except these four~\cite{Bulbul:2014sua}. A line close to this energy
is also seen towards the Andromeda galaxy
(M31)~\cite{Boyarsky:2014jta} and the center of the Milky Way
(MW)~\cite{Boyarsky:2014ska}. The measured fluxes by XMM-Newton from
Perseus (without the core), M31, and the MW are shown in
\tableref{fluxes}.  The initial analyses have motivated a great deal
of follow-up activity, including supporting evidence, null results,
and proposed explanations in terms of line emission from
ions~\cite{Riemer-Sorensen:2014yda,Jeltema:2014qfa,Malyshev:2014xqa,%
  Anderson:2014tza,Boyarsky:2014paa}.

\begin{table}[t]
\begin{center}
\caption{3.5 keV line fluxes and cored halo parameters.}
\label{table:fluxes}
\begin{tabular}{c|ccc}
\hline\hline
& Flux & $J$ & $\Sigma$ \\
& \ $10^{-6}\,\cm^{-2}\,\s^{-1}$ \ & \ $\kpc\,\gev^2/\cm^6$ \ 
& \ $\kpc\,\gev/\cm^3$ \\
\hline
Perseus & $21.4^{+7.0}_{-6.3}$~\cite{Bulbul:2014sua} & 2.3 & 20 \\
M31 & $4.9^{+1.6}_{-1.3}$~\cite{Boyarsky:2014jta} & 8.2 &  13 \\
MW & $29 \pm 5$~\cite{Boyarsky:2014ska} & 41 & 37 \\
\hline\hline
\end{tabular}
\end{center}
\vspace*{-0.2in}
\end{table}

Here we consider the possibility that this line is a signal from the
de-excitation of dark matter.\footnote{For other alternatives, see,
  for example, Ref.~\cite{Dudas:2014ixa}.}  There are two limiting
scenarios. In the ``short lifetime'' scenario, the $\gbinoX$ lifetime
is $\tau \alt 10^{15}~\s$.  $\gbinoX$ states are created in inelastic
collisions $\gbino \gbino \to \gbino \gbinoX$ and then decay.  The
signal is a dark matter analogue to the 21 cm line of neutral
hydrogen~\cite{Cline:2014eaa,Finkbeiner:2014sja,Frandsen:2014lfa} and
is proportional to $\rho^2$, where $\rho$ is the dark matter mass
density.  The predicted flux is
\begin{eqnarray}
\label{flux-short}
\lefteqn{F_{\text{short}} 
= \frac{\langle\sigma_{\gbinoX}v\rangle}{8 \pi m_X^2} 
\left \langle \textstyle{\int}
\rho^2 d\ell \right \rangle_{\text{FOV}} \text{FOV}  
= \frac{1.1 \times 10^{-3}}{\cm^2~\s}} \\ 
&\times& \! \! \! \left[\frac{\tev}{m_X}\right] 
\! \! \left[ \frac{\langle\sigma_{\gbinoX} v/c \rangle/m_X}
{10^{-3}~\text{barn}/\gev}\right] 
\! \! \left[\frac{J}{\kpc \,\gev^2 / \cm^6}\right]
\! \! \left[\frac{\text{FOV}}{\text{deg}^2}\right] , \nonumber
\end{eqnarray}
where the integral is along the line of sight, FOV is the field of
view of the measurement, $J \equiv \left \langle \int \rho^2 d\ell
\right \rangle_{\text{FOV}}$ is an average over this FOV, and
$\sigma_{\gbinoX} = \sigma (\gbino \gbino \to \gbino \gbinoX)$ is the
cross section for creating excited states $\gbinoX$.  For $m_X \sim
\tev$, the kinetic energy in $\gbino \gbino$ scattering is typically
large compared to the hyperfine splitting.  We therefore expect the
inelastic and elastic cross sections to be similar to each other and
to the transfer cross section, as is the case in an atomic dark matter
model~\cite{Cline:2013pca}.  It is tantalizing that the indicated
cross sections from self-interactions and the 3.5 keV line are roughly
similar, despite their being completely disparate phenomena.

Before we determine what values of $\sigma_{\gbinoX}$ are favored,
however, we must ask if {\em any} value of $\sigma_{\gbinoX}$ can
explain all the data.  To do this, we must determine $J$ for halo
profiles that are consistent with self-interacting dark matter and
compare them to the observed fluxes.  For the MW and M31, the FOV is a
cone with 14' half-angle. The MW observations are centered on the
galactic center.  The equilibrium self-interacting dark matter
solution~\cite{Kaplinghat:2013xca} requires that the core radius be
set by the gravitational potential of the stars, since they dominate
at the center. We use a modified NFW profile, $\rho(r)\propto
1/(r+r_c)/(r+r_s)^2$ with $r_s = 21~\kpc$ and core radius $r_c =
0.5~\kpc$, normalized to a local density of
$0.4~\gev/\cm^3$~\cite{Kaplinghat:2013xca}.  For M31, we use a similar
profile, but with a density at 8.5 kpc of $0.2~\gev/\cm^3$.  For
Perseus, we compute the flux in a projected radius of 240 kpc using
$M_{\text{vir}} = 1.1 \times 10^{15}~M_\odot$ and a concentration
parameter $R_{\text{vir}} / r_s = 6$, which gives rise to the same
surface density within 240 kpc as that in Ref.~\cite{Bulbul:2014sua} .
The resulting $J$ values are given in \tableref{fluxes}.

\eqref{flux-short} and the $J$ and flux values of \tableref{fluxes}
imply
\begin{equation}
\label{sigmashort}
\frac{\sigma_{\gbinoX}}{m_X} \sim 0.016 \ (0.005) \ [0.006] \
\frac{\text{barn}}{\gev}
\left[\frac{m_X}{\tev} \right]
\left[ \frac{\OmegaDM}{\Omega_{\gbino}} \right]^2 \! ,
\end{equation}
for Perseus (M31) [MW].  Taken at face value, these results are in
tension because we expect $\sigma_{\gbinoX}$ to follow
$\sigma_{\gbino}$ and be smaller in the clusters due to the larger
relative velocities.  At the same time, there are considerable
uncertainties from halo modeling~\cite{Finkbeiner:2014sja} and line
flux measurements.  Below, we focus on Perseus for the short lifetime
scenario, keeping this tension in mind.

Dwarf spheroidal satellite galaxies of the Milky Way constrain the
short lifetime scenario. For most dwarfs, we expect $J \agt (0.1\,
M_\odot/\text{pc}^3)^2 \times 0.6~\text{kpc} \sim 10~\kpc \,\gev^2 /
\cm^6$, using the observed commonality of halo masses within 300
pc~\cite{Strigari:2008ib}. Using $v/c \sim 10^{-4}$ and
$\sigma_{\gbinoX}/m_X = 0.01\, \text{barn}/\gev$, we predict a flux of
$2\times 10^{-6}/\cm^{-2}~\s^{-1}$, about an order of magnitude larger
than the stacked dwarf limit~\cite{Malyshev:2014xqa}.  However, for
dwarfs and $m_X \sim 100~\gev - 1~\tev$, the kinetic energy of the
collision satisfies $m_X v^2/2 \alt \Delta E$, and so a detailed
analysis of $\sigma_{\gbinoX}$ is required to predict the flux from
dwarfs.

Alternatively, in the ``long lifetime'' scenario, the $\gbinoX$
lifetime is longer than the age of the Universe, $\tau \agt
10^{18}~\s$.  The $\gbinoX$ states are created at the time of
hadronization, and since the hyperfine splitting is small compared to
the temperature at confinement, the number densities of $\gbino$ and
$\gbinoX$ are identical at that time. The $\gbinoX$ states then slowly
decay, with a signal proportional to $\rho$ and flux
\begin{eqnarray}
\label{flux-long}
\lefteqn{F_{\text{long}} = \frac{1}{4 \pi m_X \tau}
\left \langle \textstyle{\int} \rho \, d\ell 
\right \rangle_{\text{FOV}} \text{FOV}} \\
&=& \frac{7.5 \times 10^{-7}}{\cm^2~\s} 
\! \! \left[\frac{\tev}{m_X}\right] 
\! \! \left[\frac{10^{20}~\s}{\tau}\right] 
\! \! \left[\frac{\Sigma}{\kpc \,\gev/\cm^3}\right]
\! \! \left[\frac{\text{FOV}}{{\text{deg}}^2}\right], \nonumber
\end{eqnarray}
where $\Sigma = \left \langle \int \! \rho \, d\ell \right
\rangle_{\text{FOV}}$ is the surface density.  Values of $\Sigma$ for
the halo models described above are given in \tableref{fluxes}.  

\eqref{flux-long} and the $\Sigma$ and flux values of
\tableref{fluxes} imply
\begin{equation}
\label{taulong}
\tau \sim 200 \ (500) \ [300]~\text{Gyr}
\left[ \frac{\tev}{m_X} \right]
\left[ \frac{\Omega_{\gbinoX}}{\frac{1}{2} \OmegaDM} \right] ,
\end{equation}
for Perseus (M31) [MW]. Given the large systematic uncertainties in the 
M31 measurement~\cite{Boyarsky:2014jta}, these three signals
are consistent in the long lifetime scenario.

We have checked that the required lifetimes are not in conflict with
cosmic microwave background observations.  Adapting existing
constraints on the annihilation cross section of dark matter
particles~\cite{Galli:2011rz} by equating the energy injection rates
in the annihilation and decay processes at $z=1091$, we find $\tau
\agt 2~\text{Myr}\,[\tev/m_X] [2\Omega_{\gbinoX}/\OmegaDM] [\Delta
  E/3.5\,\kev]$.

\begin{figure}[t]
  \includegraphics[scale=0.48]{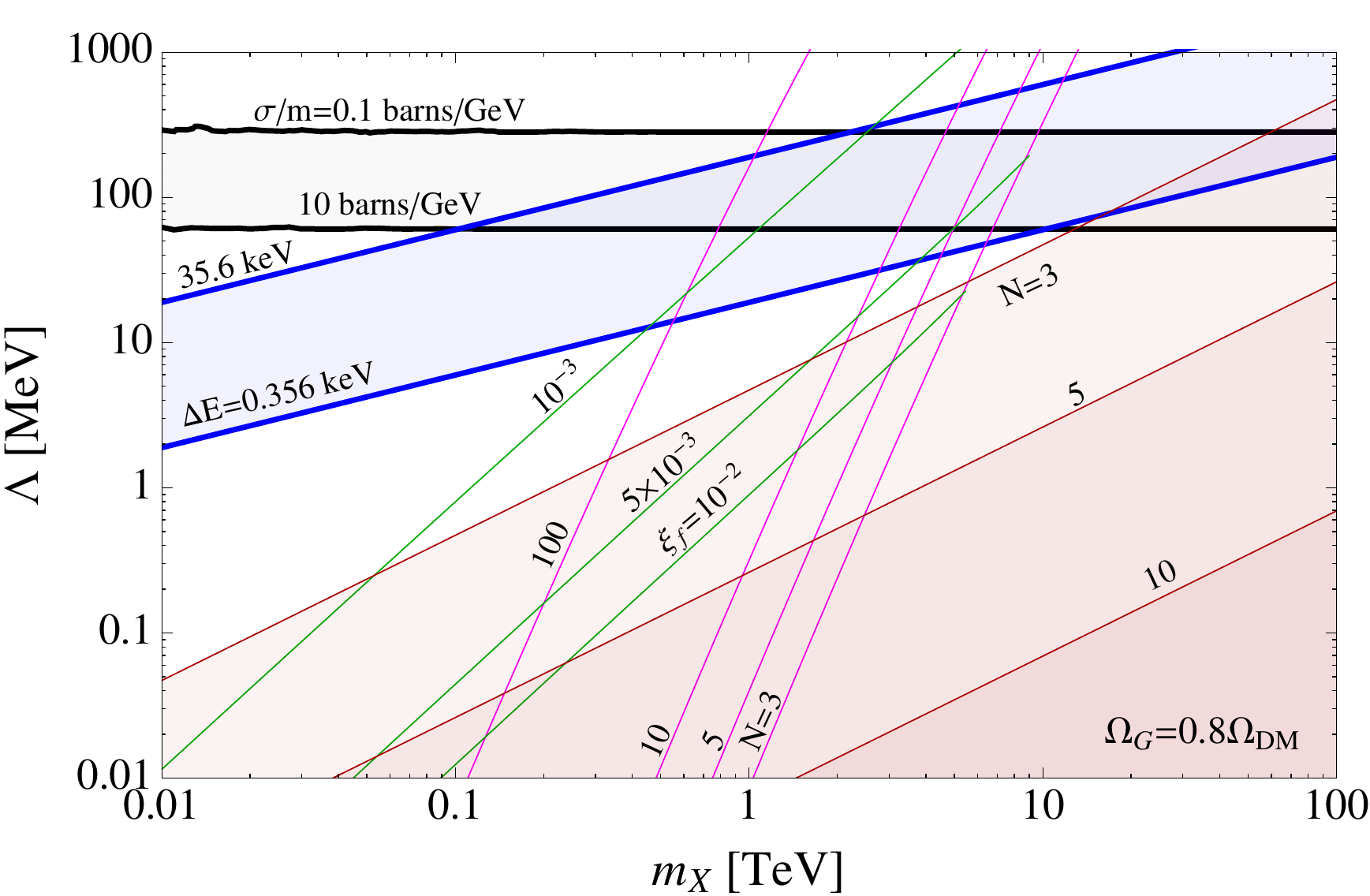}
\vspace*{-0.1in}
  \caption{Thermal SIMPle dark matter with $\Omega_{G} = 0.8\,
    \OmegaDM$ and $\Omega_{\tilde{g}} = 0.2\, \OmegaDM$.  For fixed
    $(m_X, \Lambda)$, $N$ and $\xi_f$ are determined by the relic
    densities; contours of $N = 3, 5, 10, 100$ and $\xi_f = 10^{-3}, 5
    \times 10^{-3}, 10^{-2}$ are shown.  In the indicated bands,
    $\sigma/m = 0.1-10~\text{barn}/\gev$ and $\Delta E = 0.356 -
    35.6~\kev$. Where these overlap, the model may explain both
    self-interactions and the 3.5 keV line through long-lifetime
    $\gbinoX$ decays (see text).  In the lower-right shaded regions,
    $\gbino$ re-annihilation may be significant for the values of $N$
    indicated.}
  \label{fig:long-lifetime}
\vspace*{-0.1in}
\end{figure}

\ssection{Results} We now have all the ingredients to identify viable
example models and their observational implications.  We begin by
considering a completely thermal scenario, in which the gluino and
glueball relic densities are given by
\eqsref{gbinodensity}{gbdensity}.  As an example, we consider the case
with $\Omega_G = 0.8\, \OmegaDM$ and $\Omega_{\tilde{g}} = 0.2\,
\OmegaDM$. The required values of $N$ and $\xi_f$ are shown in the
$(m_X, \Lambda)$ plane in \figref{long-lifetime}.  Relatively cold
hidden sectors are required to avoid glueballs overclosing the
Universe.

In this glueball-dominated scenario, the self-interaction cross
section is essentially $\sigma_G$, and so is in the desired range for
$\Lambda \sim 100~\mev$.  This constraint and the $\Delta E =
3.56~\kev$ band are also shown in \figref{long-lifetime}.  These bands
overlap, for example, at $(m_X, \Lambda) = (3~\tev,70~\mev)$, where
$\alpha_X \approx 0.013$, $N \approx 10$, and $\xi_f \approx 4 \times
10^{-3}$. At this point, $\sigma_{\gbinoX}$ is far too small to
explain the keV line flux in the short lifetime scenario. However, the
flux can be explained by long-lifetime decays. \Eqref{taulong} implies
$\tau \sim 30~\text{Gyr}$, which, given \eqref{lifetime}, implies a
connector mass $m_C \sim 4-6~\tev$.

\begin{figure}[t]
  \includegraphics[scale=0.48]{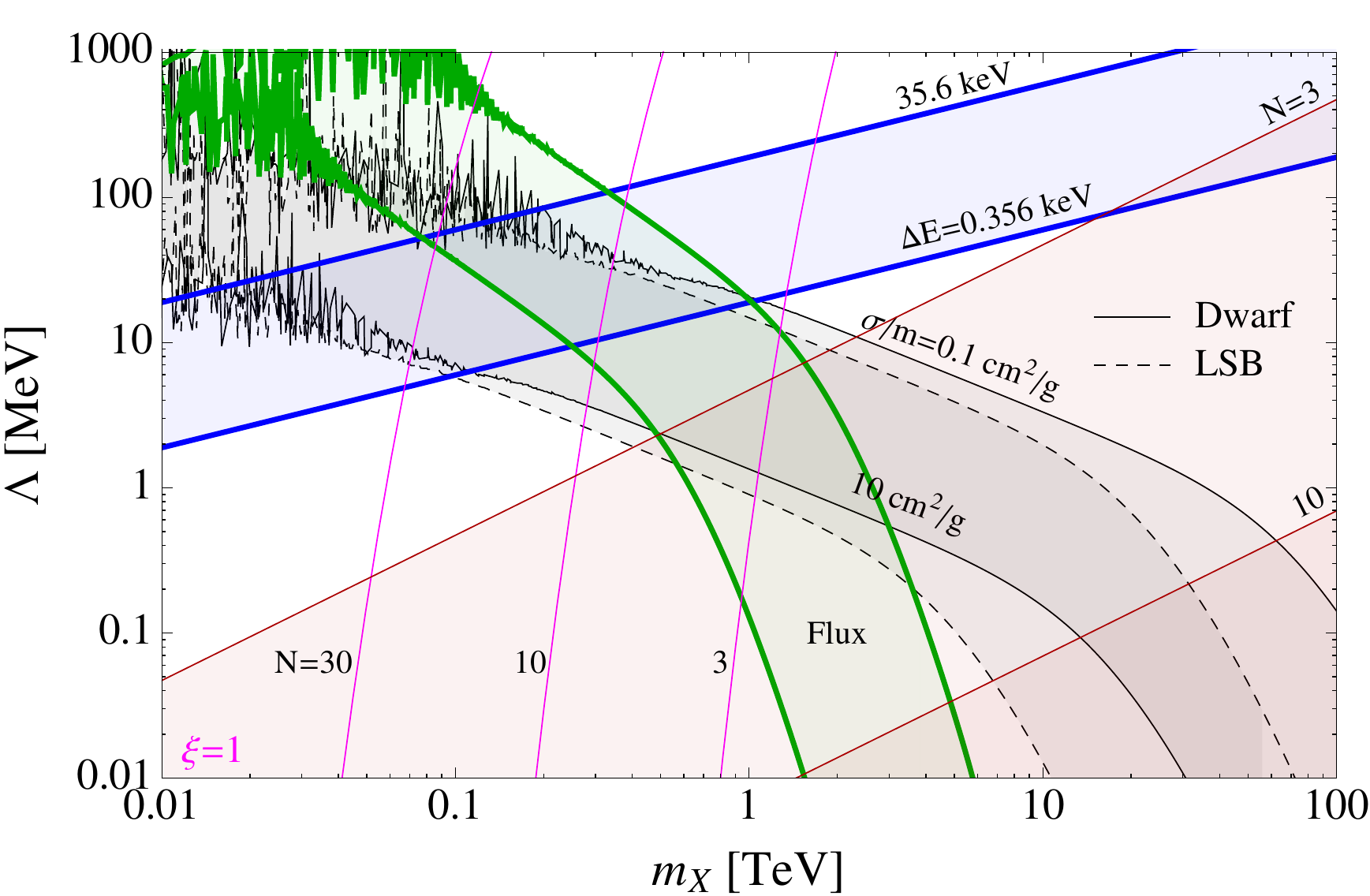}
\vspace*{-0.1in}
  \caption{Pure glueballino SIMPle dark matter with $\xi_f = 1$.  For
    fixed $(m_X, \Lambda)$, $N$ is determined by $\Omega_{\tilde{g}} =
    \OmegaDM$; contours of $N=3,10,30$ are shown. In the indicated
    bands, $\sigma/m = 0.1-10~\text{barn}/\gev$, $\Delta E = 0.356 -
    35.6~\kev$, and short-lifetime $\gbinoX$ decays give a keV line
    flux within an order of magnitude to explain the Perseus
    observations, assuming $\sigma_{\gbinoX} = \sigma_{\gbino}$.  The
    flux may also be explained by long-lifetime $\gbinoX$ decays (see
    text).  In the lower-right shaded regions, $\gbino$
    re-annihilation may be significant for the values of $N$
    indicated. }
  \label{fig:shortlifetime}
\vspace*{-0.1in}
\end{figure}

We now consider the case where the gluon density is depleted to
$\Omega_G \approx 0$ through some mechanism, such as the one of
Ref.~\cite{Boddy:2014yra} described above. Glueballs then do not
overclose the Universe for any $\xi_f$, and we consider $\xi_f = 1$.
The resulting parameters are shown in \figref{shortlifetime}.

In this pure $\gbino$ scenario, the preferred self-interactions and
keV line energy overlap, for example, at $(m_X, \Lambda) = (350~\gev,
20~\mev)$, where $\alpha_X \approx 0.019$ and $N \approx 10$. The keV
line flux may again be explained by long-lifetime decays; in this
case, \eqref{taulong} implies $\tau \sim 1000~\text{Gyr}$, which,
given \eqref{lifetime}, implies $m_C \sim 3-9~\tev$.  In this case,
however, the self-interactions also imply a large up-scattering rate,
and so the short lifetime scenario is also viable where all three
bands overlap in \figref{shortlifetime}.  A lifetime of $\tau \sim
10^{15}~\s$ implies $m_C \sim 500-700~\gev$, which is beyond collider
limits on particles that have electric charge, but not strong
interactions, in the visible sector.

\ssection{Conclusions} Currently there are tantalizing astrophysical
indications that dark matter may be self-interacting and the source of
a 3.5 keV X-ray line.  Although neither of the indications for
self-interactions and keV lines is unambiguously compelling
individually, they are both interesting, and more so if they may be
explained simultaneously in a simple model.

We have explored these in the context of a simple hidden sector: a
supersymmetric pure SU($N$) gauge theory.  The astrophysical hints
favor $m_X \sim \tev$ thermal relics interacting with $\Lambda \sim
100~\mev$ force carriers, with photons created by transitions between
highly degenerate states with $\Delta E \sim 10~\kev$.  In this model,
the qualitative hierarchy $\Delta E \ll \Lambda \ll m_X$ and the
quantitative relation $\Delta E m_X \sim \Lambda^2$ are naturally
explained by asymptotic freedom and, essentially, atomic physics.
Despite its simple formulation, the model has a rich cosmology, with
both glueballs and glueballinos contributing to dark matter, and
decays that can be either short and long compared to the age of the
Universe.  The short lifetime possibility is remarkable in that the
desired self-interactions imply a keV line flux roughly in accord with
observations, albeit with some tension between the various
datasets, while the long lifetime scenario provides a beautifully
consistent explanation for the X-ray line observed in clusters of
galaxies, M31, and MW observations.

\ssection{Note Added}
After publishing this article, we became aware of the fact that the glueball and glueballino scattering cross sections are proportional to $1/N^2$ for large $N$.
The plots present an accurate estimate of the scattering cross sections for small $N$, but the cross sections have been overestimated for large $N$.

\ssection{Acknowledgments} We are grateful to Geoff Bodwin for helpful
correspondence.  J.L.F. and Y.S.~thank the CERN Theoretical Physics
Group and J.L.F. thanks the Technion Particle Physics Center for
hospitality.  This research is supported in part by BSF Grant
No.~2010221 (J.L.F. and Y.S.), NSF Grants No.~PHY--1316792 (J.L.F. and
T.M.P.T.), No.~PHY--1214648, and No.~PHY--1316792 (M.K.), ISF Grant
No.~1367/11 and the ICORE Program of Planning and Budgeting Committee
and ISF Grant No.~1937/12 (Y.S.), DOE Grant No.~DE-SC0011632 and the
Gordon and Betty Moore Foundation through Grant No.~776 (K.B.), a
Guggenheim Foundation grant (J.L.F.), and the University of
California, Irvine through a Chancellor's Fellowship (T.M.P.T.).

\bibliography{bibxrayline}

\end{document}